\documentclass[twocolumn]{aa} 

\usepackage{graphicx}
\usepackage{subcaption}
\usepackage{xcolor}
\usepackage{soul}
\usepackage{txfonts}
\usepackage{hyperref}

\usepackage{booktabs}
\usepackage{multirow}
\usepackage[T1]{fontenc}
\usepackage{tabularx,ragged2e,multirow}
\newcolumntype{C}{>{\Centering}X}

\begin{document}

   \title{Deciphering transmission spectra by exploring the solar paradigm}
   
   \author{N.-E. Nèmec,
          \inst{1,2} Ò. Porqueras-León\inst{1,2}, I. Ribas\inst{1,2}
          , and A. I. Shapiro\inst{3,4}}

   \institute{Institut de Ciències de l’Espai (ICE, CSIC), Campus UAB, c/ Can Magrans s/n, 08193 Bellaterra, Barcelona, Spain,\email{nnemec@ice.csic.es} 
   \and
   Institut d’Estudis Espacials de Catalunya (IEEC), Edifici RDIT, Campus UPC, 08860 Castelldefels (Barcelona), Spain
\and Institut f\"ur Physik, Universit\"at Graz, Universit\"atsplatz 5/II, 
   8010 Graz, Austria, 
\and Max-Planck-Institut f\"ur Sonnensystemforschung \\ Justus-von-Liebig-Weg 3, 37077 G\"ottingen, Germany}

  \abstract
   {Transmission spectroscopy allows measuring the wavelength dependence of the transit signal of an exoplanet, thus enabling probing its atmospheric composition. However, the transmission spectrum also carries information of the host star, generally referred to as 'contamination'. Stellar activity leads to an apparent change in the stellar radius, thus, directly impacting the transit depth. This contamination is regarded as the major hurdle in discovering and characterising atmospheres of exoplanets.}
   {The objective is to understand how the chromatic effect (i.e. the wavelength dependence) of the stellar activity-induced  transit depth depends on the surface distribution of magnetic features. The surface distribution of other stars generally is unknown, with the exception of our very own star, the Sun. We therefore investigate the solar paradigm as 'ground-truth' to explore how much the chromatic effects depends on the distribution of magnetic features. In particular, we explore the impact of centre-to-limb variations (CLV) of the magnetic features and their resulting chromatic effect. 
   Specifically, we investigate the solar paradigm as the 'ground truth'.}
   {We utilise spot and faculae masks obtained from SDO/HMI magnetograms and intensitygrams together with the SATIRE approach of calculating solar variability to calculate the chromatic dependence of the apparent radius  of the Sun for the last solar cycle. We test several approaches in convolving the area coverages with the spectra to uncover the potential biases, and investigate the drivers responsible for the chromatic effect.}
   {We find that using a simplified approach that only relies on the disc area coverage and neglects CLV in the spectra to calculate the chromatic effects lead to an underestimation of the apparent radius. In particular, for the faculae component the CLV need to be taken into account accordingly, especially since the facular area coverage is by far larger than that of spots for stars with near-solar activity level.  We report that this chromatic dependence can be detected in transits of an Earth- and Jupiter-sized planet. We additionally assessed the amplitude of this effect between solar minimum and solar maximum. We found that for a Jupiter-like transit this amplitude of 40 ppm, well above the 10 ppm noise-floor of JWST. However, the effect is only on the 0.4 ppm level for the Earth-like transit. }
   {}


\maketitle

\section{Introduction}

Exoplanet research has advanced tremendously since the first detection of an exoplanet around a star similar to our Sun \citep{Mayor1995}. Not only methods that allow for the detection of small rocky planets have been developed, but in parallel techniques for the characterisation of exoplanetary atmospheres have been established. The most powerful technique to study and characterise the atmospheres of exoplanets at present is transmission spectroscopy \citep{Seager2000, Charbonneau2002, Tinetti2007, Yan2019, Rustamakulov2023}. The planet will appear larger at wavelengths where the atmosphere is more opaque, which, in turn, depends on its chemical composition. Thus, by studying the dependence of the depth of the planetary transit on the wavelength (called transmission spectrum) information on the abundances of relevant chemical compounds can be
obtained. On the other hand, the star's apparent radius can also vary with wavelength \cite{Seager2024}, particularly if it exhibits magnetic activity, which manifests as dark or bright regions on its surface \citet[see][for a review on magnetically driven stellar activity]{Solanki2013}.\\
Technical efforts to detect and characterise Earth-like planets have pushed the capabilities of our instruments to the detection limit of such planets, and the remaining hurdle is the stellar signal \citep{Rackham2023}. While signal itself provides precious information of the star, its activity in particular, it is often seen as 'noise' plaguing efforts of retrieving planetary characteristics. The stellar activity signal has different sources, depending on the timescales considered. Most prominently it is the stellar magnetic activity on cool stars, like our Sun, manifesting itself in in the form of bright and dark regions(generally referred to as spot and faculae) on stellar surfaces \citep{Solanki2013, Ermolli2013}. Whilst spot or faculae crossings can be seen in the individual transit lightcurves \citep[for instance][]{PetitditdelaRoche2024, Ahrer2025, Fournier-Tondreau2025,Radica2025}, un-occulted features pose further threats to the reliable characterisation of planetary atmospheres.

While there are some promising attempts underway to correct for stellar activity contamination from the transmission spectrum (i.e. leveraging contemporaneous multi-technique observations \cite{Mallonn2018, Rosich2020}, or machine learning approaches \cite{Ardevol2022, Matchev2022,Lueber2024}), we still largely rely on forward models of stellar activity. In such models, a rotating stellar photosphere is filled with magnetic features (spots and faculae) \citep[see for instance][for a general overview of the approach]{Rackham2019}. Usually, a grid of models, with varying active region coverages and configurations, as well as temperature contrasts of the features with respect to the quiet photosphere, is generated. Then, iteratively, models from such a grid are fitted to the observed wavelength-dependent lightcurves. After the best fit model is found, this chromatic stellar model is removed from the observed transmission spectrum to remove the stellar contamination \citep[f.i.][]{Lim2023, Radica2025}. The stellar radius, or rather the star-to-planet radius ratio is often also considered a free parameters in the fitting. In short, forward models rely on two components: the spectra of magnetic features and the quiet stellar surface, and their spatial distribution of the surface of the star. A lot of effort has been put into improving the fidelity of of spectra for the magnetic activity components \citep{Witzke2022}. \cite{Rosich2020} have shown that the same projected area coverage, but different configurations of magnetic features lead to different contaminations in the transmission spectrum.
This raises the question, how well constrained the magnetic feature distribution has to be in order to remove the contamination, or if there is some leeway.
As the distribution of magnetic features on other stars is unknown, we choose to focus on the only star for which this is known: the Sun.
To this end, we utilise the observed distribution of magnetic features of the Sun by using spot and facula masks obtained from SDO/HMI images and magnetograms. These features are then weighted  with their corresponding spectra taken from \cite{Unruh1999}. We then follow the approach of \cite{Seager2024} to calculate the apparent radius of the star, in our case the Sun, and compare this to a star that would be free of magnetic features. This way, we model the effect that un-occulted features have on the transmission spectrum of a planet. We introduce the model in more detail in Sec. \ref{Model}.  The apparent radius change and its causes are discussed in Sec. \ref{Results}. We conclude the paper and its is limitations to use for other stars in Sec. \ref{Summary}.

\section{Model}\label{Model}

The transit depth, $TD(\lambda)$ induced by a transiting planet generally is written as

\begin{equation}
    TD(\lambda) = \frac{R(\lambda)_{p}^2}{R(\lambda)_{\star}^2},
\label{eq:apparent change}
\end{equation}

\noindent with $R(\lambda)_{p}$ being the radius of the planet and $R(\lambda)_{\star}$ the apparent radius of the star. The apparent size of the star is varying with time due to the presence of magnetic features; in the presence of a large spot, the stellar radius appears to 'shrink' (the spot temperature is cooler than the non-magnetic photosphere, hence the measure effective temperature is cooler than the photospheric effective temperature). Therefore, we rewrite $R(\lambda)_{\star}^2$ as $R(\lambda)_{\star, A}^2$  with $A$ denoting a star with some level of magnetic activity.
We stress once again that the apparent radius in Eq. \ref{eq:apparent change} is only due to the presence of stellar activity, and not a physical change in the radius of the star (as for instance oscillations would induce).

Following \cite{Seager2024}, and taking into account that we are only concerned with the Sun in the following, we use $\odot$ instead of $\star$, and write the apparent solar radius $R(\lambda)_{\odot, A}^2$ as 
\begin{equation}
    \pi R(\lambda)^2_{\odot, A} = \pi R^2_{\odot, Q} \left ( 1- \sum_i \alpha_i \left( 1 - \frac{F(\lambda)_i}{F(\lambda)_Q}\right)\right), 
\label{eq:radius_change}
\end{equation}

\noindent where $R^2_{\odot, A}$ denotes the apparent solar radius in the presence of magnetic features on the surface, $R^2_{\odot, Q}$ denotes the 'quiet' (i.e. a Sun that is free of magnetic features) solar radius,  $\alpha_i$ is the fractional coverage by the considered magnetic feature $i$ (either spot or facula), $F(\lambda)_i$ is the flux of the given feature, and $F(\lambda)_Q$ is the quiet Sun flux.
As discussed in Seager et al. (2024) the approach in Eq. 2 does not take into account centre-to-limb variations (CLV) in the spectra, and the area coverage $\alpha_i$ represents the disc area coverage, neglecting the actual distribution of the features.\\
Our 'ground-truth' approach utilises the observed distribution of the magnetic features (in the present study faculae and spots), which is obtained following \cite{Yeo2014} using images from the Helioseismic and Magnetic Imager onboard the Solar Dynamics Observatory \citep[SDO/HMI][]{Schou2012} spanning from 2010 May to 2021 January. Following the Spectral And Total Irradiance REconstruction \citep[SATIRE][]{Fligge2000, Krivova2003} approach, we introduce the facular saturation threshold (which determines the area coverage from spot corrected HMI magnetograms), $B_{sat}$. $B_{sat}$ is the only free parameter in the model. Since the magnetic elements that form faculae are too small to be resolved in the full disk HMI magnetograms, we assume that the facular filling factor increases linearly with increasing magnetic flux up to a saturation point, $B_{sat}$, above which the pixel is assumed to be filled fully by faculae and the filling factor is one \citep{Ball2012}.
We use $B_{sat}$ = 305 G as discussed in \cite{Kopp2024}. We utilise one image per day (if available), and each image represents a 12 minute average. This gives us enough coverage to capture the day-to-day variability of the solar activity, which is primarily driven by magnetic activity \citep{Solanki2013, Ermolli2013,Shapiro2017}.
We therefore rewrite Eq. \ref{eq:radius_change} to take into account  the fractional area by spots and faculae and their position-dependent flux  in each pixel of HMI images with coordinates $m,n$

\begin{equation}
    \pi R(\lambda)^2_{\odot, A} = \pi R(\lambda)^2_{\odot,Q} \left ( 1- \sum_{m,n}\sum_i \alpha_i^{mn} \left( 1 - \frac{F(\lambda)_i^{mn}}{F(\lambda)_{\odot,Q}^{mn}}\right)\right).
\label{eq:radius_change_mu}
\end{equation}

\noindent  For each of the magnetic components, we utilise the commonly used 1D fluxes for quiet Sun, spot (divided in umbra and penumbra) and faculae computed
by \cite{Unruh1999} \citep[following][]{Castelli1994} with the use of the spectral synthesis code ATLAS9 \citep{Kurucz1992}. To take CLV into account, those fluxes are calculated at 10 different $\mu$-positions on the solar disc, where $\mu$ $=$ $\cos(\theta)$ ($\theta$ is the angle between the observer's direction and the position vector defined with respect to the centre of the Sun). 
We limit ourselves to the wavelength regime from 0.6 -- 6 $\mu$m, covering the wavelength regimes of the JWST NIRSpec \citep{Jakobsen2022, Boeker2022} instrument, while preserving the original resolution of the spectra.

To investigate the importance of both the distribution of magnetic features and their contrasts, we utilise the previously introduced calculations of the apparent radius and denote the approaches in the following way

\begin{itemize}
    \item Approach 1: Utilising Eq. \ref{eq:radius_change} and calculating the mean over the $\mu$-positions  of the given spectra, i.e. $F(\lambda$) $=$ $\int F (\lambda,\mu) d\mu$ $F(\lambda)$ $=$ $\langle F (\lambda,\mu) \rangle_{disc}$
    \item Approach 2: Utilising Eq. \ref{eq:radius_change_mu}, representing the 'ground-truth' in the present work
\end{itemize}

\noindent We remark that the calculations presented below correspond to chromatic dependencies that would occur in the transmission spectrum of a planet that does not cross any magnetic feature, i.e. the transmission spectrum contaminated by \textit{un-occulted} features. The scaling relations presented later in this work therefore can be used as a template to correct for un-occulted features in transmission spectra.

The outlined approaches for calculating the activity-induced apparent stellar radius due to stellar activity differ in the type of spectra considered. Approach 1, using $F(\lambda$) $=$ $\int F (\lambda,\mu) d\mu$,  corresponds to a solid-angle weighted average disc integrated spectrum. In Approach 2, which we treat in the following as the ground truth, represents the most accurate description of the apparent solar radius, by taking into account the distribution of the magnetic features and their respective spectra properly.

\section{Results}\label{Results}

\subsection{Activity induced solar apparent radius change}

\begin{figure*}[t]
\centering
\includegraphics[width=\linewidth]{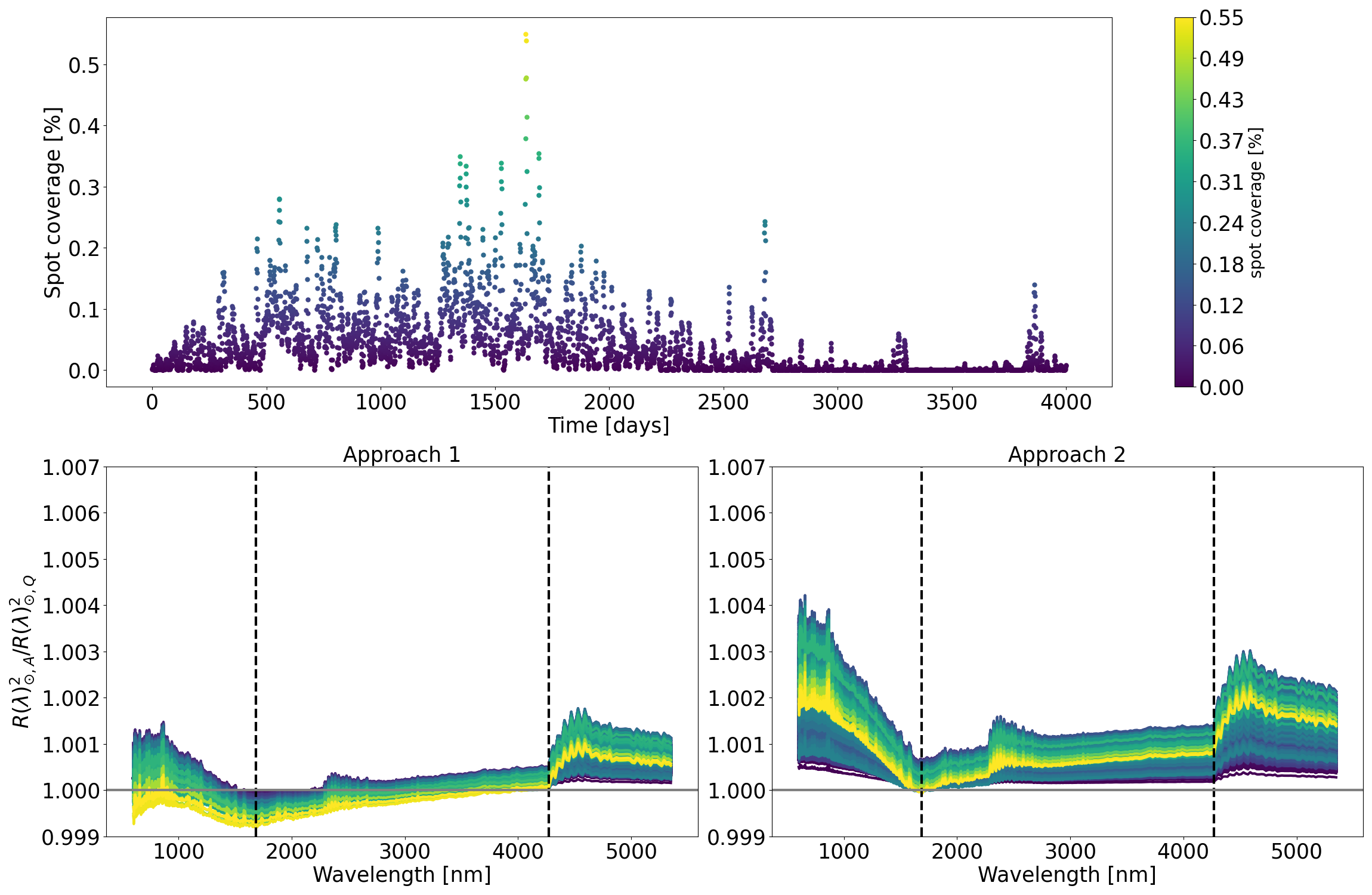}
\caption{Dependence of the apparent solar radius as a function of solar activity for the different approaches. Top figure shows the spot area coverage as a function of the time for solar cycle 24. The colours in the bottom row plots correspond to the spot disc area coverage. Approach 1 corresponds to the disc integrated spectra and Approach 2 takes the spectra of each pixel of the HMI maps into account.}
\label{fig:apparent_radius_change}
\end{figure*}

\begin{figure}
\centering
\includegraphics[width=0.45\linewidth]{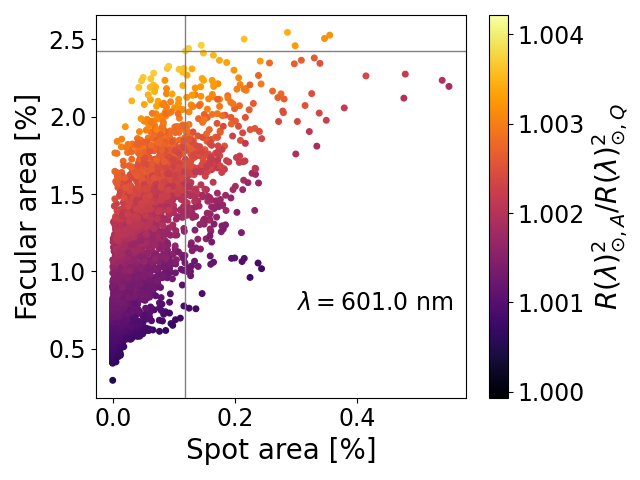}
\includegraphics[width=0.45\linewidth]{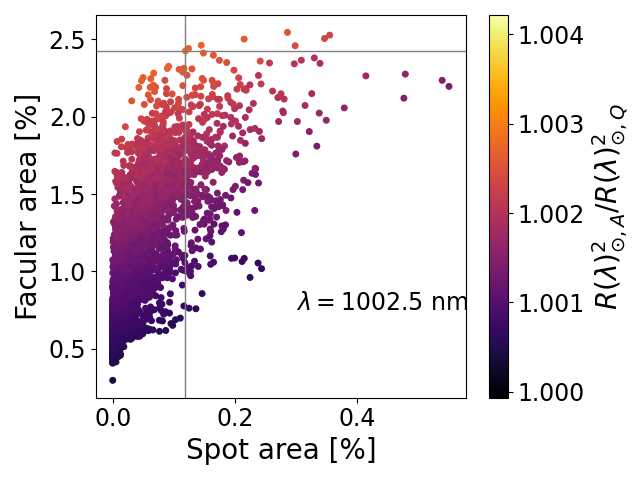}
\includegraphics[width=0.45\linewidth]{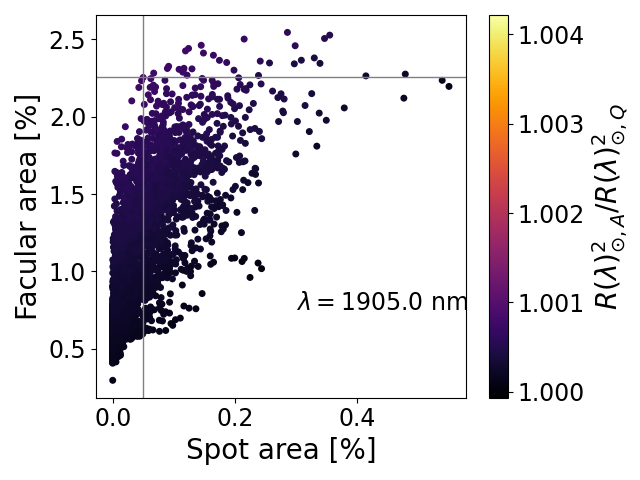}
\includegraphics[width=0.45\linewidth]{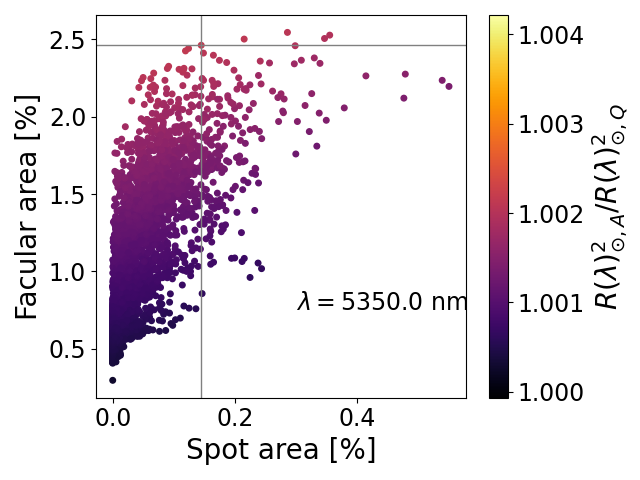}
\caption{Activity induced apparent radius (as indicated by the colours) as a function of spot area coverage for selected wavelengths. The grey lines indicate the position of the maximum in the apparent radius for each of the wavelengths shown. We note, that the average facular area is 1\% and the mean spot area is 0.04\%.} An animated version of this figure is available as Supplementary Material.
\label{fig:filling_factors_radius_change}
\end{figure}

In Fig. \ref{fig:apparent_radius_change} we show the results of the calculations with the different approaches outlined above. The top panel displays the spot area coverage (our chosen activity indicator) as a function of time for solar cycle 24. The spot areas represent the disc area coverage obtained from the HMI maps. The same goes for the faculae area coverages. We remind the reader that faculae and spot area coverages are connected, albeit not tightly correlated, with the faculae area generally increasing less rapidly with increasing spot area coverage \citep[see i.e.][and Fig. \ref{fig:filling_factors_radius_change} in the present work]{Chapman1997,Nemec_faculae}. 
The bottom panels in Fig. \ref{fig:apparent_radius_change} show the apparent radius calculated with the two approaches. Already a first glance at the apparent radius for the different approaches reveals major differences. Approach 1 shows a clear uniform trend from the active Sun (green and yellow colours corresponding to high spot area coverages near solar maximum), to the less active Sun, whereas in Approach 2 most of the curves are overlapping, showing no clear trends with activity level. 
Quantitatively,  Fig. \ref{fig:apparent_radius_change} shows that Approach 1 underestimates the apparent radius  at medium to high activity levels compared to the ground truth represented via Approach 2. The curves representing Approach 1 show a simple shift between the chromatic dependencies below 4200 nm, but a decrease in the slopes for wavelengths below. Approach 2 on the other and shows the shift below 1600 nm, but there is a difference between the curves in the tilt below 1600 nm. We will discuss this in more detail later in this work. 
In important implication from Fig. \ref{fig:apparent_radius_change} and Approach 2 (the ground truth) is that unnocculted faculae cannot explain the downward slopes in transmission spectra reported in the literature. 
In Fig. \ref{fig:filling_factors_radius_change} we show, in addition to the dependence of the facular area on the spot area, also how the apparent radius (indicated via the colorbar) is depending on said areas. For the wavelengths shown, we conclude that the largest apparent radius can be found at high facular, but intermediate spot areas.

\begin{figure*}
\centering
\includegraphics[width=0.80\linewidth]
{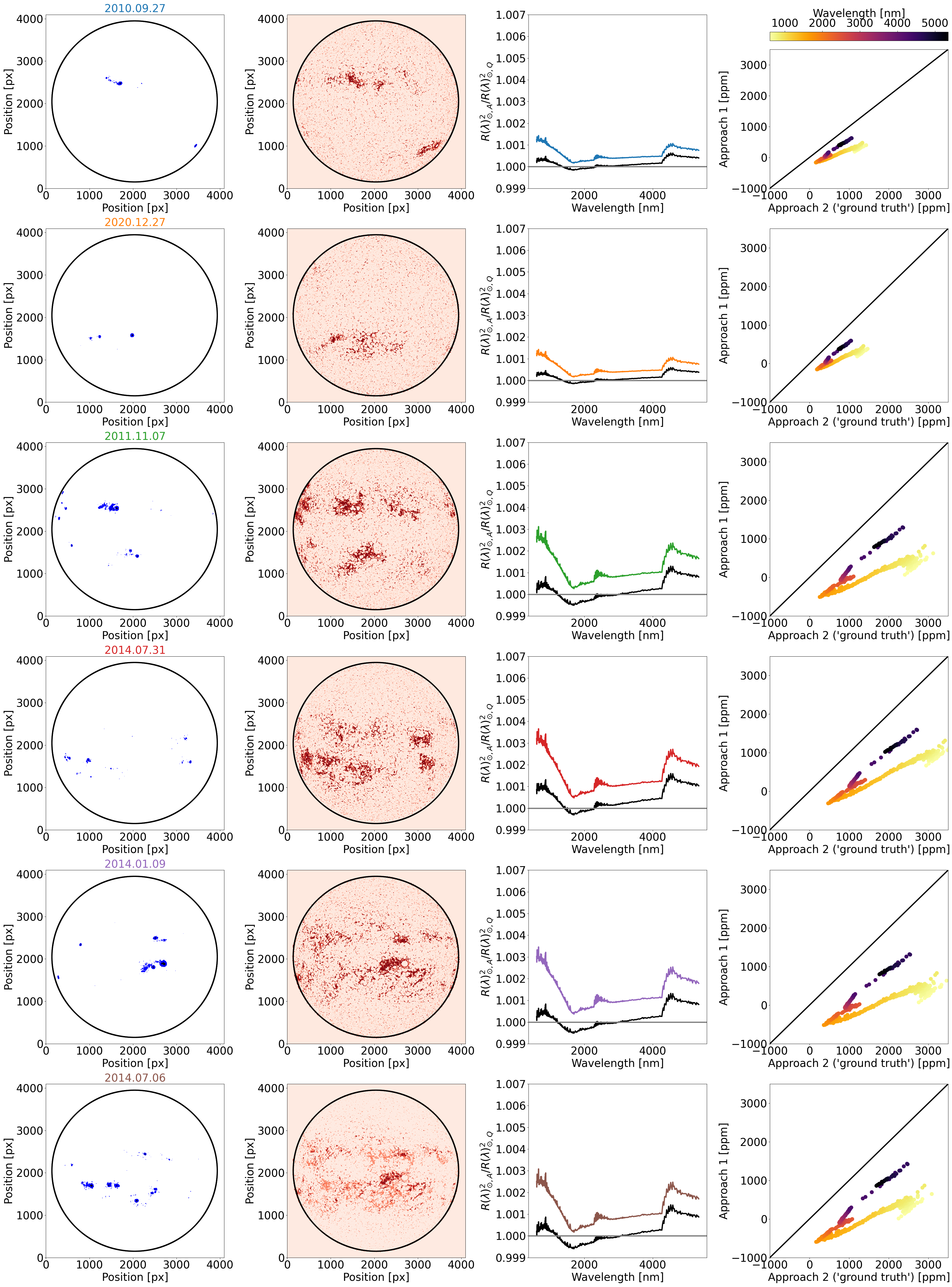}
\caption{A closer look at the different approaches presented in Fig. \ref{fig:apparent_radius_change} by taking six days presented in Tab. \ref{tab:areas}, with their spot distribution (first column) and the faculae distribution (second column),  the apparent radius for Approach 1 (black solid lines) and Approach 2 (coloured solid lines) (column three), and scatter-plots of the apparent radius as a function of wavelength (column four). For details of why those dates were chosen we refer to the text.}
\label{fig:maps}
\end{figure*}

To better study the differences, we selected six different days of solar cycle 24. Tab. \ref{tab:areas} gives the spot and facular areas for those dates. We note, that the dates were chosen so that the area coverages are almost the same (row 1 and 2, row 3 and 4 and row 5 and 6) and as a result also the facula-to-spot area ratio, but their actual distribution on the solar surface is different. 
We show in Fig. \ref{fig:maps} the resulting chromatic dependences calculated with Approach 1 (black lines) and 2 (coloured lines) and also the corresponding actual spot and faculae area distributions taken from HMI images. We find that none of the curves for Approach 1 and 2 for the chosen dates overlap. For the days of low solar activity (2010.09.27 and 2020.12.27), the curves are the closest, however Approach 1 still underestimates the apparent radius by around 200 ppm. The scatter plots in the 4th column of Fig. \ref{fig:maps} show the differences more clearly. On the x-axis we plot Approach 2, y-axis gives Approach 1 and the coloured lines indicate the wavelength (lighter colours indicating shorter wavelengths), with the black solid lines indicate a 1-1 correspondence between the two approaches. Interestingly, the apparent radius as a function of wavelength of Approach 1 as a function of Approach 2, shows two distinct branches, splitting at around 2500 nm. The scatterplots further offer evidence that Approach 1 always underestimates the apparent radius and that the functional form of the apparent radius is sensitive to the treatment of the CLV. What is also interesting, is that 4 dates (2011.11.07, 2014.07.31, 2014.01.09, 2014.07.06) have very different spot area coverages, but similar facular areas, resulting in different ratios. Interestingly, the resulting apparent radius change is the same. This shows, again, that overall the faculae are more dominant over the spots. We will come back to this point later in the paper.

\begin{figure}
\centering
\includegraphics[width=\linewidth]{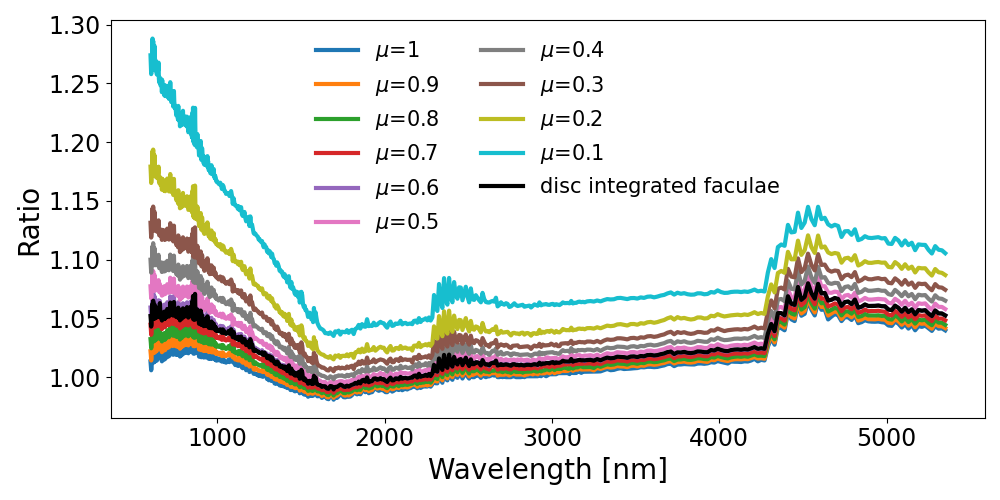}
\caption{Flux ratio between the faculae and the quiet Sun for the different approaches. In black line Approach 1 (black line) is shown and the coloured lines indicate the flux rations for different $\mu$-positions .}
\label{fig:contrasts_faculae_mu}
\end{figure}

\begin{figure}
\centering
\includegraphics[width=\linewidth]{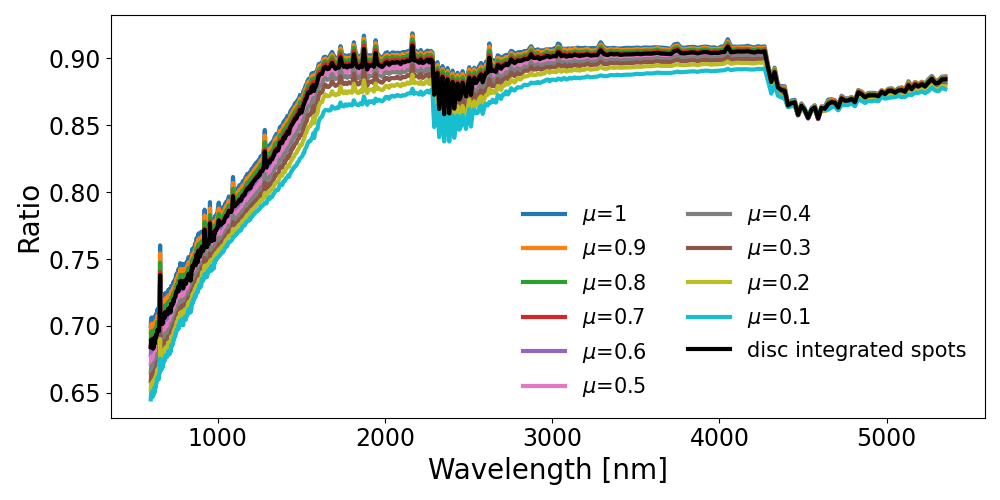}
\caption{Similar to Fig. \ref{fig:contrasts_faculae_mu}, but for the spots.}
\label{fig:contrasts_spots_mu}
\end{figure}

We have found that using a disc averaged spectrum to calculate the chromatic apparent radius through Approach 1  underestimates the activity-induced radius. To understand this in more detail, we need to further analyse the contrasts of the magnetic features with respect to the quiet star, or rather, the ratio of the fluxes present in Eqs. \ref{eq:radius_change}  and \ref{eq:radius_change_mu} Fig. \ref{fig:contrasts_faculae_mu} shows the ratios between the fluxes at different $\mu$-values for the faculae and Fig. \ref{fig:contrasts_spots_mu} for the spots. We combined the umbra and penumbra contribution for the spot contrast,  by adding the umbra and penumbra spectra in a 1:5 ratio, which is representative of the average observed sunspot ratio \citep{Brandt1990}. In both figures, the black line represents the contrast that is given by the $\mu$-averaged spectrum, as used in Approach 1. The faculae fluxes are highly dependent on the disc-position, with the contrast increasing towards to limb, whereas the spot fluxes show a comparatively lower dependence.

In Approach 1, the contrasts shown by the black lines are simply multiplied by the unweighted disc area coverage. The spot ratio to the quiet solar surface is generally larger, and even the generally larger faculae areas cannot outweigh this dimming at high activity levels in Approach 1.
 Approach 2 complicates this picture, as each pixel from the HMI images is weighted with the according spectra for each $\mu$-values. Hence, the CLV of the faculae becomes very important due to their distribution. Noteworthy is that the faculae contrast on the Sun is the brightest towards the limb. This limb-brightening is pivotal to take into account in any forward modelling approaches of the variability of Sun-like stars. The limb-brightening also is crucial for the shape of the chromatic dependence of the apparent radius, particularly below 1600 nm. This wavelength regime corresponds to the opacity minimum in the solar atmosphere, where photons are formed in the deepest layers. In particular, due to the physical nature of the faculae, we can look into deeper layers of the atmosphere, where temperatures are lower than in the surrounding, quiet Sun layers, which are at higher atmospheric heights and higher temperatures. This temperature contrast leads to a negative brightness contrast in the faculae and in turn to the dip in the apparent radius. 
 
What is counter-intuitive is that the apparent radius is larger than unity towards the longer wavelengths, where we would expect the spots to take over. But faculae are dominating in area coverage over the spots (see Fig. \ref{fig:filling_factors_radius_change}), and their contrast with respect to the immaculate solar photosphere is still net positive, whereas the contrast of the spots to the immaculate surface decreases much faster. This means that it cannot be ruled out that faculae are still contributing at this longer wavelengths for Sun-like stars, and specifically, Sun-like stars that are less active than the Sun. We therefore stress that the treatment of faculae is  crucial for stars with similar activity levels than the Sun at wavelengths in the infrared, but even more so for stars with lower activity levels \citep{Nemec_faculae}.

\begin{table}
\caption{The disc area coverages of spots and faculae, as well as the facula-to-spot ratio from the maps shown in Fig. \ref{fig:maps}}
\begin{tabular}{cccc}
\hline
\textbf{Date} & \textbf{Spot area [\%]} & \textbf{Facular area [\%]} & Ratio \\
\hline \hline
2010.09.27 &  0.06 & 0.89 & 14.8  \\
2020.12.27 &  0.06 & 0.83 & 13.8  \\
2011.11.07 & 0.28 & 2.03  & 22.5 \\
2014.07.31 &  0.09 & 2.02 & 23   \\
2014.01.09 & 0.31 & 2.07 &  6.67  \\
2014.07.06 &  0.31 & 2.36 & 7.6  \\
\hline
\end{tabular}
\label{tab:areas}
\end{table}

\begin{figure}
\centering
\includegraphics[width=\linewidth]{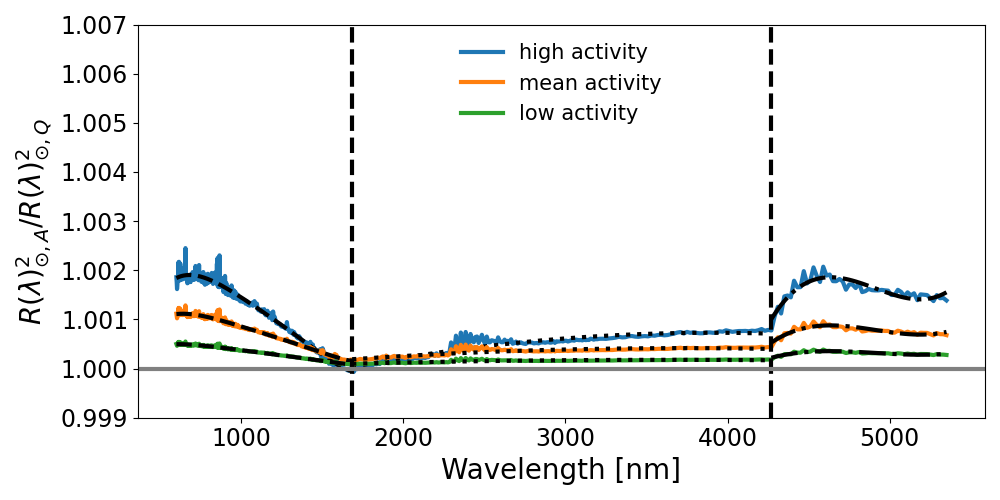}
\caption{Example of arbitrarily chosen spectra and fits to different wavelength
regimes. The dashed black lines correspond to the fit given in Eq.
\ref{eq:fit1}, the dotted black line to Eq. \ref{eq:fit2}, and the black dash-dotted line to Eq. \ref{eq:fit3}}
\label{fig:fits_example}
\end{figure}

We want to quantify the chromatic dependence of the apparent radius as a function of activity, to provide scaling relations for forward models and retrieval approaches.
As the chromatic effect is rather complex, we divide the spectra into three different parts, according to wavelength regime (the limits are shown in the bottom row of Fig.\ref{fig:apparent_radius_change}), and for each regime we utilise a different functional form for the fit, aiming for maximising the correlation coefficient in finding the best fit:

\begin{align} 
\frac{R(\lambda)_{\odot,A}}{R(\lambda)_{\odot,Q}} &=  a\lambda^5 + b\lambda^4 + c \lambda^3 + d\lambda^2 + e\lambda + f \quad \textrm{if} \quad\lambda \leq \textrm{1600 nm} \label{eq:fit1}\\ 
\frac{R(\lambda)_{\odot,A}}{R(\lambda)_{\odot,Q}} &=  a\lambda + b \quad \textrm{if} \quad\textrm{1600 nm} \leq \lambda \leq \textrm{4200 nm} \label{eq:fit2}\\ 
\frac{R(\lambda)_{\odot,A}}{R(\lambda)_{\odot,Q}} &=  a\lambda^4 + b\lambda^3 + c \lambda^2 + d\lambda + e \quad \textrm{if} \quad \lambda \geq \textrm{4200 nm} \label{eq:fit3}
\end{align}

In the $\lambda \le 1600$\,nm regime, we note that a second-order polynomial fit in principle yielded similar R-squared values as the fifth-degree polynomial one chosen, but the former did not capture the tilt in the dependence below 700 nm. The chromatic dependence between 1600--4200 nm is almost flat with the exception of the feature at around 2500 nm which is a feature of water , which we do not take into account in our functional form, for simplicity. The 4200 nm cut-off was chosen to capture the CO-feature at around 4500 nm. In Fig. \ref{fig:fits_example} we show spectra chosen for different activity levels as a demonstration of the fits. The activity levels correspond to the facula and spot values given in Tab. \ref{tab:coefficients}.

   \begin{table*}
     \caption{Coefficients of the fits to the chromatic dependencies shown in Fig. \ref{fig:fits_example}. The activity levels were chosen the following: "High activity" was taken to be the day of highest active region coverage (0.55\% spots and 2.2\% faculae) , "mean activity" to the day with the mean spot area coverage area coverages (0.04\% for the spots and 0.81\% for the faculae), and the "low activity " case is represented via the day of the minimum faculae area coverage of 0.28\% with no spots present on the solar disc at all.}
     \label{tab:coefficients}
     \centering
     \begin{tabular}[width=0.5\textwidth]{cccccccc}
\hline
      & \textbf{a} & \textbf{b} & \textbf{c} & \textbf{d} & \textbf{e} & \textbf{f} & \\ 
        \hline   \hline
    \textbf{\textless 1600 nm} &  &  &  &  &  &  & \\ 
    high activity  & 2.52321e$-$18 & $-$1.50460e$-$14 & 3.55279e$-$11 &$-$4.1637e$-$08& 2.31909e$-$05& 0.99638\\
    mean activity & 
    2.43605e$-$18& $-$1.45248e$-$14& 3.42851e$-$11 &$-$4.0156505e$-$08 &2.23412e$-$05& 0.99653\\
    low activity & 2.21247e$-$18 &$-$1.31913e$-$14& 3.11365e$-$11 &$-$3.64672e$-$08& 2.028801e$-$05&  0.99685 \\ 
      \hline
    \textbf{1600 $-$ 4200 nm} &  &  &  &  &  &  & \\ 
    high activity  & 8.39718e$-$08 &1.00015&  $-$ & $-$ &  $-$&   $-$ & \\ 
    mean activity & 8.018170e$-$08& 1.00015&$-$ & $-$ & $-$ &  $-$  & \\ 
    low activity & 7.28585e$-$08 &1.000136 & $-$ & $-$ & $-$ & $-$ & \\
      \hline
    \textbf{\textgreater 4200 nm} &  &  &  &  &  &  & \\ 
    high activity &  $-$4.11559e$-$15 &8.12944e$-$11 &$-$6.013729e$-$07& 0.0019743& $-$1.42584 & $-$ & \\
    mean activity &$-$3.96901e$-$15& 7.83974e$-$11& $-$5.79931e$-$07& 0.0019039& $-$1.33926& $-$& \\ 
    low activity  &$-$3.60320e$-$15& 7.11722e$-$11& $-$5.264871e$-$07& 0.001728& $-$1.12372 &$-$&\\
    \hline
     \end{tabular}
   \end{table*}

\begin{figure}
\centering
\includegraphics[width=\linewidth]{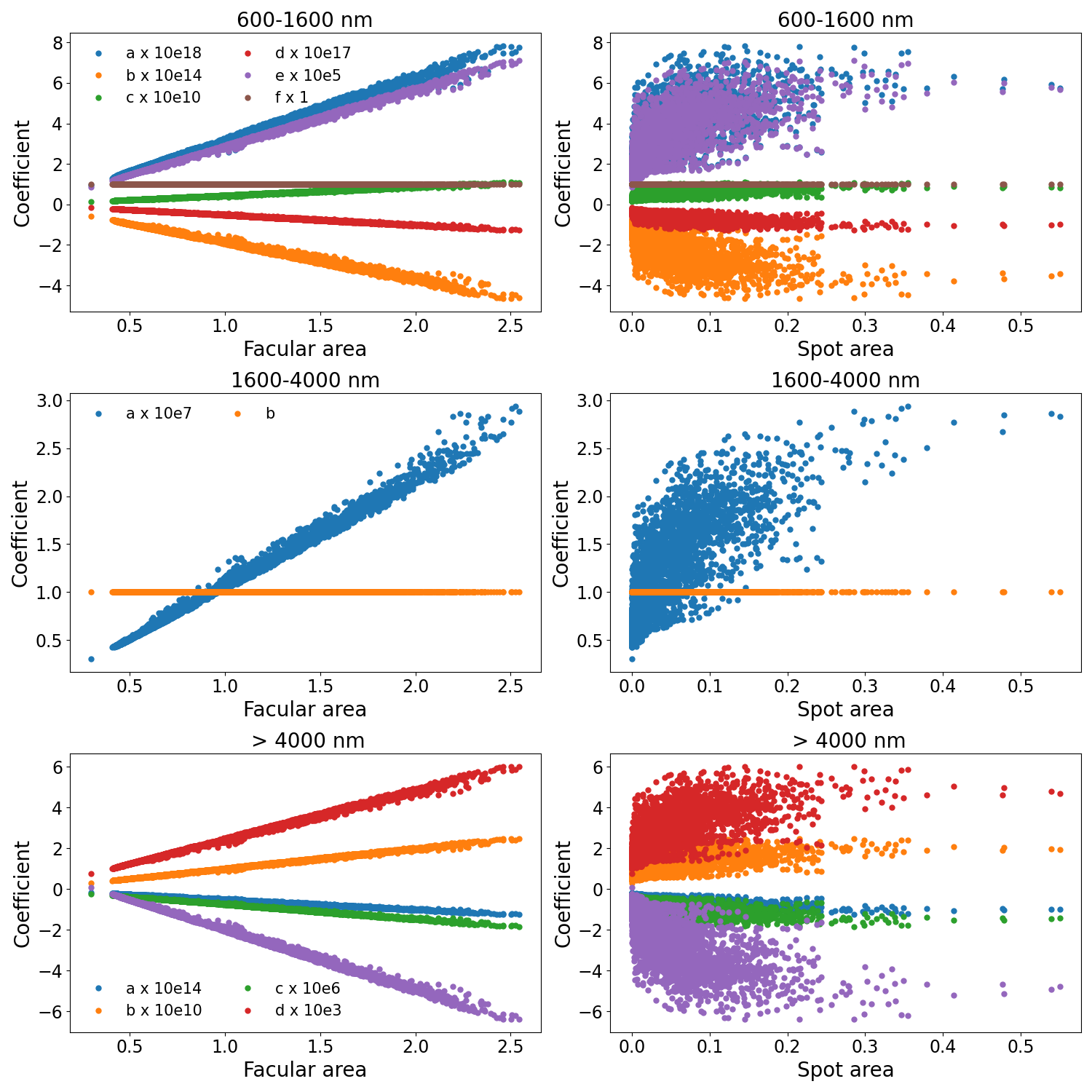}
\caption{Dependence of the fit coefficients for the different wavelength regimes from Eqs. \ref{eq:fit1}--\ref{eq:fit3} on the faculae and spot area coverages. The coefficients were multiplied by different factors as given in the legend in order to put them on equivalent scales.}
\label{fig:fits_coefficients}
\end{figure}

We provide the coefficients of the fits for the three different activity levels as indicated in Fig. \ref{fig:apparent_radius_change} in Tab. \ref{tab:coefficients}. Since the apparent radius as a function of wavelength is orders of magnitudes smaller than the wavelength regimes used for the fitting, the coefficients are generally found to be very small, but spanning various magnitudes. We also investigated the dependence of the fit coefficients on the spot and faculae area coverages separately (shown in Fig. \ref{fig:fits_coefficients}), and found that the coefficients all linearly depend on the faculae area coverage. This is not surprising, as the faculae are the dominant driver of the brightness variations on the Sun. The coefficients show a parabolic dependence on the spot area and generally there is more scatter (compared to faculae). The parabolic trend is a direct consequence of dependence of the facular on spot-area shown in Fig. \ref{fig:filling_factors_radius_change}. This figure also explains the scatter with the spot area. Additionally, faculae are longer-lived on the Sun, whereas most spots live less than one solar rotation. Generally, while there is a correlation between facular and spot area coverages, as shown in Fig. \ref{fig:filling_factors_radius_change}, this correlation is not a tight one \citep[see][for further discussion]{Chapman1997, Foukal1998, Nemec_faculae}.

\subsection{Impact on the transit depth and connection to JWST}

\begin{figure}
\centering
\includegraphics[width=\linewidth]{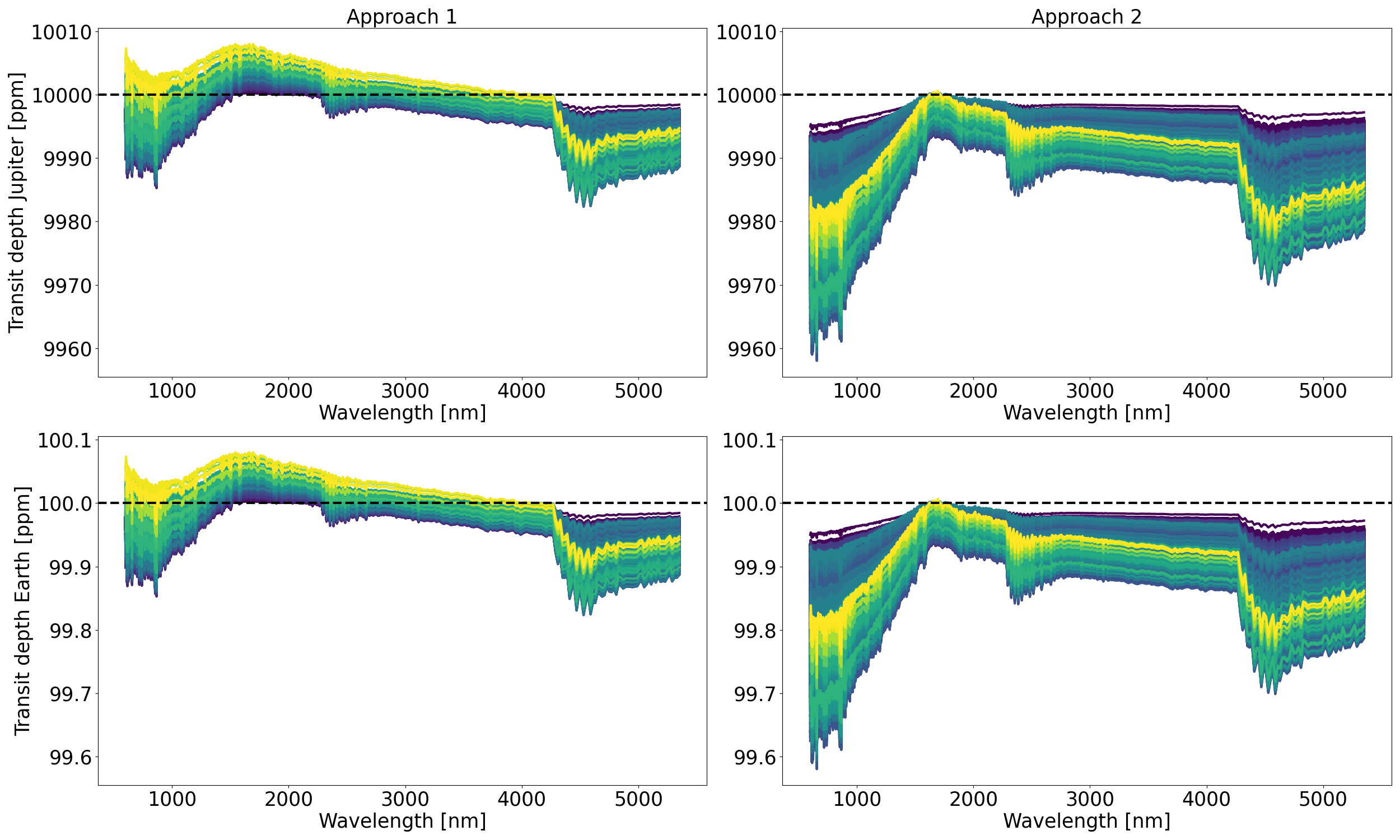}
\caption{Chromatic dependence of the transit depth for a Jupiter-like (top) and Earth-like bottom planet. The dashed lines indicate the transit depth that is a result of the difference in radii with respect to the Sun (hence in the absence of both stellar activity and a planetary atmosphere).}
\label{fig:transit}
\end{figure}

\begin{figure}
\centering
\includegraphics[width=0.65\linewidth]{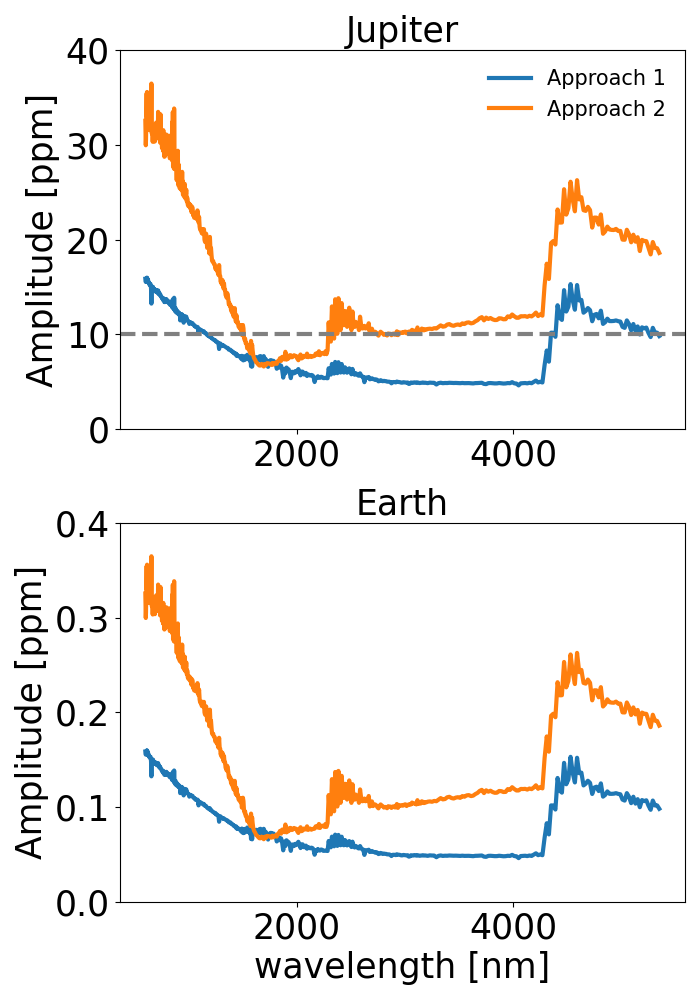}
\caption{Amplitude of the impact on the transit depth induced by solar activity (i.e. the difference between solar maximum and minimum. The grey dashed line on the top panel is the 10 ppm noise floor level for JWST as found by \cite{Rustamakulov2023}.}
\label{fig:amplitude}
\end{figure}

We have discussed in the previous section the impact of the solar activity on the apparent radius. Lastly we now want to consider the impact of the apparent solar radius on the transit depth of a planet itself. For this, we utilise Eq. \ref{eq:radius_change} and the solar radius as calculate and presented in Fig. \ref{fig:apparent_radius_change}. We consider cases of two transiting planets, a jupiter - and an earth-sized one. The transit depth of a Jupiter-like planet around the Sun is 10 000 ppm, for the Earth it would be 100 ppm, solely based on the radii compared to the Sun. We also assume no chromatic effect of the planets (i.e. a planet without an atmosphere), we want to consider the impact of the solar apparent radius solely. The results of these calculations are presented in Fig. \ref{fig:transit}. Unsurprisingly, towards the visual part of the spectrum, the slope is positive and the transit depth lower. At around 16000 nm, in Approach 2, the planet radius is almost at the assumed 10 000 and 100 ppm level, respectively, as the apparent solar radius in the presence of solar activity is at a minimum (i.e. there is no change from the 'quiet', activity free solar radius.). For Approach 1 around 1600 nm, the transit becomes deeper due to the spots dominating the chromatic dependence.
\cite{Rustamakulov2023} estimated the noise floor of JWST to be around 10 ppm. We therefore want to now look at the amplitude of signal that the change in the level of the solar activity induces during the two transits considered. For this, we take the curves from Fig. \ref{fig:transit} and calculate the minimum and maximum value of the transit depth for each wavelengths (i.e. the minimum and maximum activity induced contamination of the signal). We show the ratios of the minimum and maximum signal (hence the amplitude of the effect) in Fig.\ref{fig:amplitude}. The amplitude of the change in the solar level of activity is too small to be detected for an Earth like transit (the values are below 10 ppm, however for a jupiter-like transit in the 'ground-truth' model (Approach 2) the amplitude is high enough to be detected by JWST, hence presents a hurdle for the retrieval of a planetary atmosphere. We emphasise, that this only means that the change in the amplitude of the apparent radius between low-and high solar activity cannot measured by JWST, however the transit depth of an Earth-like (as seen in Fig.\ref{fig:transit}) is above the noise-floor.

\section{Summary and Outlook}\label{Summary}

A planet's transit depth depends, at first order, on the squares of the planet-to-star radius ratio. As stellar activity in the form of spots and faculae introduces a change in the apparent radius, it is crucial to understand the chromatic effect of the stellar variability in order to correct for it in atmospheric retrieval methods \citep{Lim2023}.
 We investigated two approaches to calculate the chromatic dependence of the apparent radius by using the Sun as a testbed, as it is the only star for which the active region distribution is  well constrained. We found that calculating the apparent radius by the using simply the disc area coverage of a given feature and their disc integrated spectrum \citep[i.e., as used by][]{Seager2024}, generally underestimates the apparent radius compared to using the sophisticated approach that takes foreshortening effects and CLV into account. This underestimation would mean that the stellar contamination is not correctly removed in the transmission spectrum, and hence the planetary radius will be overestimated. 
We parametrised the chromatic dependence of the apparent radius and have shown quantitatively that the faculae component is the important driver behind changes in the flux, and is in particular important for wavelengths below 2000 nm. This is unsurprising as observations of both the Sun \citep[for instance][]{Chapman1997, Foukal1998, Shapiro2016, Nemec_faculae} and solar analogues \citep[i.e.][and references therein]{Lockwood2007,Radick2018} show that the the faculae contribution to stellar brightness variations is dominating over the spot contribution for low activity stars like the Sun. We also presented fit coefficients for different activity levels for a Sun-like star as a recipe for mitigating stellar activity. We emphasise, that the fitting coefficients presented are only applicable to stars with similar distributions in terms of the latitude of active regions (i.e. as depicted by the butterfly diagram) to the Sun.

Additionally, our results clearly show that on stars with similar activity than the Sun, un-occulted faculae will dominate the chromatic radius dependence of the star. In particular, if not taken into account properly in removing the stellar contamination in the transmission spectrum, the radius of the planet will be overestimated greatly. So far in the literature, cooler stars than the Sun have been targeted, hence it is not straightforward to apply our findings to those stars. In particular, JWST is mostly observing K- and M-type dwarfs, where the star-planet-ratio is larger (than for instance for the Sun-Earth system). Recent  developments in spectral modelling \citep[i.e.][]{Norris2023}, indicate that faculae on M-dwarfs look very different than on the Sun, namely that their brightness contrast is negative. This puts claims of the literature for stars like TRAPPIST-1 showing signs of un-occulted faculae \citep[i.e][]{Lim2023} under scrutiny. Our results however can be used for forward-modelling exercises for the upcoming Ariel mission \citet{Tinetti2022}.

The important implication of our results is that the negative slope in the transmission spectra in the visual wavelength regime as reported in the literature can be partially explained by un-occulted faculae. However, two things have to be kept in mind. Firstly, often retrieval methods do not account for CLV effects \citep[for instance][]{PetitditdelaRoche2024, Radica2025}. Secondly, faculae are approximated with a spectrum that is hotter than the stellar photosphere. As shown by \cite{Witzke2022} this is not true for the Sun even. In particular, a hotter stellar atmosphere, would not allow for the observed limb-brightening by faculae, which also can be seen in simulations of faculae on other spectral types \cite{Norris2023}. This particular highlights the importance of proper adoption of faculae models in retrieval codes, and the need for better synergies between the exoplanet, stellar astrophysics and solar physics community. 

We have employed 1D models of both spots and faculae in the present work,  where in particular the limb darkening (or rather brightening) for the faculae was adjusted for the solar case.  \cite{Witzke2022} have shown that the 1D stellar atmosphere models that are regularly used to model the stellar contamination spectra do not agree with novel faculae models that are based on the 3D MDH Max Planck Institute for Solar System Research (MPS)/University of Chicago
Radiative Magnetohydrodynamics \citep[MURaM,][]{Voegler2004, Rempel2014} simulations. In particular, the '3D' faculae show an even stronger negative contrast around 1500 nm than predicted by the 1D models \citep{Witzke2022} This once more highlights the need of physics-based models of stellar activity, especially for the faculae. Qualitatively, for our results that would mean, that the slope in the visible wavelength regime is steeper. Additionally, at around 1500 nm the apparent radius might reach values slightly below unity, as the faculae on the 3D simulations show a stronger negative contrast than their 1D counterparts.

Despite the limitations of current approaches commonly found in the literature, our results clearly show, that for a star with the Sun, the change in the activity level cannot currently be detected by JWST if we consider an Earth-like planet, as the effect is below the noise floor for its instruments. However, in the case of a Jupiter-like transit, stellar activity contamination is easily detectable and therefore should be properly accounted for, especially in the case of un-occulted features.

The question of how to mitigate the stellar contamination if the magnetic field distribution is unknown, which is the case for any other star than the Sun, still remains. In particular, we have shown that the CLV plays a large role, and that same distributions read to the same results. The reality however is that even filling factors (i.e. area coverages) themselves are hard to constrain, and are treated as free parameters in forward models \citep[i.e.][]{Lim2023, Radica2025} to mitigate stellar activity in exoplanet atmosphere retrievals. As we have shown in the present work, there are degeneracies due do the exact distribution of the features, if the filling factors are very similar.
The question still remains how we can estimate the filling factors. \cite{Mallonn2018, Rosich2020} have shown that it is possible to constrain filling factors and surface distributions better by utilising contemporaneous multi-technique (i.e. high-resolution spectroscopy combined with multi-colour photometry). In a follow up work, we are planning to revisit such an approach and utilise available solar data (i.e. SSI - SORCE, HARPS-N spectral information) and investigate, with the Sun as ground truth once again, the potential of using stellar activity indicators to mitigate the stellar contamination in transmission spectroscopy.

\begin{acknowledgements}

We thank the anonymous referee for their comments that helped improve the manuscript.
This publication has been made possible by the spanish grants PID2021-125627OB-C31 funded by MCIU/AEI/10.13039/501100011033 and by “ERDF A way of making Europe”, PID2020-120375GB-I00 funded by MCIU/AEI, by the programme Unidad de Excelencia María de Maeztu CEX2020-001058-M, by the Generalitat de Catalunya/CERCA programme, by the SGR 01526/2021, the European Research Council (ERC) under the European Union’s Horizon Europe  programme (ERC Advanced Grant SPOTLESS; no. 101140786,  ERC Synergy Grant REVEaL grant No. 101118581) and the Marie Sk\l{}odowska-Curie Actions grant agreement No 101149286 (INCITE).

\end{acknowledgements}

\newpage

\bibliography{bib}
\bibliographystyle{aasjournal}

\end{document}